\documentclass[aps,prb,twocolumn,showpacs]{revtex4}  
\usepackage{epsfig}
\begin{document}
 \title{Intrinsic spin current for an arbitrary Hamiltonian and scattering potential}
 \author{Alexander  Khaetskii}
\address{Institute of Microelectronics Technology, Russian Academy of Sciences, 142432, Chernogolovka, Moscow District, Russia}

\date{\today}

\draft

\begin{abstract}
We have described  electron spin dynamics in the presence of the spin-orbit interaction and disorder using the spin-density matrix method. Exact solution  is obtained for an arbitrary 2D spin-orbit Hamiltonian and arbitrary smoothness of the  disorder potential. Spin current depends explicitely on the disorder properties, namely the smoothness of the disorder potential, even in the ballistic limit when  broadening by scattering is much smaller than the spin-orbit related splitting of the energy spectrum. In this sense universal intrinsic spin current does not exist.
\end{abstract}

\maketitle

Spin-orbit coupling brings about a number of  interesting effects, one of which   is generation of a spin flux in the plane perpendicular to the charge current direction. This phenomenon occurs in the paramagnetic system and has been  very well known for quite a long time, see Ref.\cite{Dyakonov}. It results from the fact that in the presense of spin-orbit coupling the scattering by impurities has  an asymmetric character \cite{Landau}. 
Extrinsic contribution exists only beyond the Born approximation in the scattering amplitude and leads to an accumulation of the spin density near the sample surface \cite{Dyakonov}. 

 It has been recently claimed \cite{MacDonald,Zhang} that an analogous phenomenon can exist even without scattering by impurities, i.e. in the ballistic regime, the corresponding contribution being called intrinsic.  Later several papers appeared where the effect of scattering  by impurities  was taken into account for the Rashba model 
 \cite{Bauer,Halperin,Khaetskii,Raimondi,Dimitrova,Chalaev}, besides, the extrinsic contribution was  described \cite{Halperin1,DasSarma} to explain the recent experiment done by Awschalom's group\cite{Awschalom}, 
 see also experiment \cite{Wunderlich}. 
\par
Here we study the dependence of the intrinsic contribution to the spin current on the form of the Hamiltonian and properties of the disorder potential using the spin-density matrix method \cite{Dyakonov1}. 
We have considered an arbitrary Hamiltonian, for example, the generalized  2D Rashba Hamiltonian with an 
arbitrary momentum dependence of the spin-orbit term or the 3D Luttinger Hamiltonian,  and the arbitrary smoothness of the disorder potential. In the case of the generalized 2D Rashba model and the arbitrary smoothness of the disorder potential an exact solution for spin current in the ballistic limit  $\Delta \tau \gg 1$  is obtained, where $\Delta$ is the spin splitting of the electron spectrum and $\tau$ is the transport scattering time. Spin current depends explicitely on the disorder properties, namely, the smoothness of the disorder potential, even in the ballistic limit. 
In this sense  universal intrinsic spin current does not exist. 
The problem of 2D holes with the $p^3$ spin-orbit term has been numerically considered recently by several groups \cite{Nomura,Sheng}.  Analytically it was studied in Ref.\onlinecite{Loss1} for the $\delta$-scattering,
 where the incorrect answer was obtained due to the wrong calculation method.  
 The case of the arbitrary smoothness of the disorder potential has been analytically  considered very recently 
 in Refs.\cite{Shytov,Shytov2,Khaetskii2}, some results were  also reported in Ref.\cite{Khaetskii1}. 
 Below we will compare the results obtained here with those from Ref.\onlinecite{Shytov2}. I also refer the reader to the introductory part of that paper where a complete list of references on spin current is presented.

\par

We consider the generalized 2D Rashba model which can be applied, for instance,  for the case of 2D holes. 
The Hamiltonian of the problem is
\begin{equation}
\hat {\cal H({\bf p})}=\frac{p^2}{2m}+ \frac{\alpha p^N}{2}\vec{\sigma}\cdot\vec{
\Omega}({\bf p}),
\,\,\,  \epsilon_M(p)= \frac{p^2}{2m}+ M\alpha p^N,
\end{equation}
where $\alpha$ is the spin-orbit coupling constant, $\Omega_x =\cos(N\phi)$, $\Omega_y =\sin(N\phi)$, 
$\Omega_z=0$, $\epsilon_M(p)$ the eigenvalues, $ M=\pm 1/2$  the helicity values. The eigenfunctions are
$$
\chi_{M{\bf p}}= \sum_{\mu=\pm 1/2} D^{(1/2)}_{\mu M}(\vec{\Omega})u_{\mu}=
 \sum_{\mu=\pm 1/2}e^{-i\mu N\phi}d^{(1/2)}_{\mu M}(\frac{\pi}{2})u_{\mu},
$$
where $D^{(1/2)}_{\mu M}(\vec{\Omega})$ is the rotation matrix \cite{Landau}, 
 $\phi$  the angle of ${\bf p}$, $N$ the winding number, and
$u_{\mu}$  the eigenfunction of the $\hat \sigma_z$ operator.
\par

We will calculate the $q_{yz}$ component of the spin current. This quantity is defined as
\begin{widetext}
\begin{equation}
q_{yz}=\sigma_s E_x = Tr \int \frac{d^2p}{(2\pi)^2} \hat f({\bf p})\frac{1}{2} (\hat S_z\hat V_y +
\hat V_y \hat S_z)= -\frac{1}{2}\int \frac{d^2p}{(2\pi)^2} 
\frac{p_y}{m}(f_{+-}({\bf p}) + f_{-+}({\bf p})).
\label{q}
\end{equation}
\end{widetext}
Here $\hat f({\bf p})$ is the Wigner spin density matrix, $\hat V_y$ the $y$-component of the velocity operator and $\hat S_z=(1/2) \hat \sigma_z$  the spin operator. The last expression in  Eq.(\ref{q})
 is given in the helicity basis. The different components of the spin density matrix have the following relations to the average spin components: 
\begin{eqnarray}
<S_z> \propto (f_{+-}+f_{-+}),\,\,\, \nonumber \\
 <\vec{S}\cdot[\vec{n} \cdot \vec{\Omega}]> \propto 
 (f_{+-}-f_{-+}), \,\,\, \nonumber \\
 <\vec{S}\cdot\vec{\Omega}> \propto 
 (f_{++}-f_{--}).
\label{relation}
\end{eqnarray}
where ${\bf n}$ is the unit vector normal to the 2D plane (z-axis). 
\par
 The general expression for the quantum kinetic equation in the case of spin-orbit interaction, when the Hamiltonian and the Wigner distribution function are matrices over the spin indexes, was derived, for example, in Refs.\cite{Dyakonov1,Shytov}.
 In our  case  this equation is simple and reads
\begin{equation}
\frac{\partial \hat f({\bf p})}{\partial t} + e{\bf E}\frac{\partial \hat f({\bf p})}{\partial {\bf p}} + \frac{i}{\hbar}[\hat {\cal H({\bf p})}, \hat f] = St\{ \hat f({\bf p})\}
\label{kinetic}
\end{equation} 
Here $e$ is the charge of the carriers. The last term on the left hand side is the commutator and the expression for the collision term is given below. Now we write Eq.(\ref{kinetic}) in the helicity basis where the Hamiltonian  is diagonal. When doing that, we should take into account the fact that eigenfunctions  $\chi_{M{\bf p}}$ depend on the direction of the momentum ${\bf p}$, thus the matrix elements of the derivative $\partial \hat f/\partial {\bf p}$ in this basis do not coincide with the quantities $\partial  f_{MM'}/\partial {\bf p}$
$$
\left(\frac{\partial \hat f}{\partial {\bf p}}\right)_{MM'} =  
\frac{\partial  f_{MM'}}{\partial {\bf p}} - \frac{i}{\hbar} [\hat {\bf a}, \hat f]_{MM'}; \,\, {\bf a}_{MM'} = i \hbar \chi^{\star}_{M{\bf p}}\frac{\partial \chi_{M'{\bf p}}}{\partial {\bf p}}.
$$
 We see that there  appears  the commutator of the vector matrix $\hat {\bf a}$ with $\hat f$.
Thus for Eq.(\ref{kinetic}) in the linear response regime (${\bf E} \parallel x$)  we obtain 
\begin{widetext}
\begin{equation}
eE \cos\phi \frac{\partial f^{(0)}_{MM}}{\partial p}
\delta_{MM'}-\frac{iN}{2}\frac{\sin \phi}{p}eE ( f^{(0)}_{M'M'}(p)- f^{(0)}_{MM}(p) ) +
\frac{i}{\hbar}(\epsilon_M(p)-\epsilon_{M'}(p))f_{MM'}({\bf p})= St (\hat f({\bf p}))_{MM'}
\label{kinetic1}
\end{equation}
\end{widetext}
Here $f^{(0)}_{MM}(p)$ is the equilibrium Fermi function corresponding to the helicity value $M$.
The collision term  in the helicity basis has the form \cite{Koshelev,Shytov}
\begin{widetext}
\begin{eqnarray}
St (\hat f({\bf p}))_{MM'}= 
 \int\frac{d^2 {\bf p}_1}{(2\pi \hbar)^2} \sum_{M_1M'_1}
\{[\delta(\epsilon_{M_1}(p_1)-\epsilon_{M}(p)) +\delta(\epsilon_{M'_1}(p_1)-\epsilon_{M'}(p))]K^{MM'}_{M_1M'_1}(\omega_{{\bf p}{\bf p}_1}) \cdot f_{M_1M'_1}
({\bf p}_1)- \nonumber \\
 -\delta(\epsilon_{M_1}(p)-\epsilon_{M'_1}(p_1))[K^{MM_1}_{M'_1 M'_1 }(\omega_{{\bf p}{\bf p}_1}) \cdot f_{M_1M'}({\bf p}) + f_{MM_1}({\bf p})\cdot K^{M_1M'}
_{M'_1 M'_1}(\omega_{{\bf p}{\bf p}_1}) ] \}, 
\label{St}
\end{eqnarray} 
\end{widetext}
where the kernel in the Born approximation in the scattering amplitude is: 
\begin{equation}
K^{MM'}_{M_1M'_1}(\omega_{{\bf p}{\bf p}_1})= \frac{\pi}{\hbar}
|U({\bf p} -{\bf p_1})|^2 \cdot n_i D^{(1/2)}_{MM_1}(\omega_{{\bf p}{\bf p}_1})
D^{(1/2)\star}_{M'M'_1}(\omega_{{\bf p}{\bf p}_1}).
\label{kernel}
\end{equation}
Here $n_i$ is the 2D impurity density, $U({\bf p} -{\bf p_1})$ is the Fourier component of the impurity potential.  
The quantities $D^{(1/2)}_{MM_1}(\omega_{{\bf p}{\bf p}_1})$ depend only on the scattering angle $\theta =\phi -\phi_1$:
\begin{eqnarray}
D^{(1/2)}_{1/2,1/2}= D^{(1/2)}_{-1/2,-1/2}= \cos (N\theta/2), \nonumber \\ 
D^{(1/2)}_{1/2,-1/2}= D^{(1/2)}_{-1/2,1/2}= -i \sin (N\theta/2).
\label{D}
\end{eqnarray} 
In this paper we consider only the case of small spin-orbit splitting, $\Delta/E_F \ll 1$, $\Delta=\alpha p^N_F$, and
will be keeping only the terms linear in this small parameter. Note that expanding Eq.(\ref{St}) with respect to this parameter,  for the collision term  in spin basis we obtain the expression which is identical to that given by Eq.23 of ref.\onlinecite{Shytov}. We will be solving the problem in the helicity basis which is much more convenient and the solution can be obtained much easier. Moreover,  further we will consider only the ballistic case,  $\Delta \gg \hbar/\tau$, when one expects the intrinsic value for the spin current. In the case when the impurity potential has the azimuthal symmetry, it is possible to obtain a solution for the scattering potential with an arbitrary correlation length $R$. The most simple cases correspond to the following situations: 
1) $ m \tilde\alpha /\hbar \ll 1/R$, 2)  $1/R \ll m \tilde\alpha /\hbar $, $\tilde\alpha = \alpha p_F^{N-1}$, 
$\hbar/m\tilde\alpha$ is the spin-orbit length. 
We will see that these two cases correspond to different answers for the spin current. 
The first case means a relatively short-ranged potential and was considered in Refs.\onlinecite{Shytov,Shytov2}. 
Note that it includes the limit of small angle scattering when $ m \tilde\alpha /\hbar \ll 1/R \ll k_F$. 
\par
As here we consider  the ballistic case, $\Delta \gg \hbar/\tau$, the problem can be greately simplified, 
namely, in the collision term, Eq.(\ref{St}), we can neglect all the nondiagonal components of the spin density matrix
since only diagonal components can be proportional to $\tau$, which follows from Eq.(\ref{kinetic1}).  
Moreover, since on the left hand side of Eq.(\ref{kinetic1}) there are only the first harmonics of the angle, 
the following  solution obeys the system of Eqs.(\ref{kinetic1},{\ref{St}): 
\begin{equation}
 f_{++}({\bf p})= f_{++}(p)\cos\phi, \,\, f_{--}({\bf p})=f_{--}(p)\cos\phi, 
\label{angle1} 
\end{equation}
\begin{equation}
 f_{+-}({\bf p})= f_{+-}(p)\sin\phi, \,\, f_{-+}({\bf p})= f_{-+}(p)\sin\phi.
\label{angle2}
\end{equation}
It can be immediately seen from Eq.(\ref{St}) with the help of the fact that the scattering kernel  
\begin{eqnarray} 
W(\theta)= \frac{n_i}{2\hbar^3}|U(\theta)|^2,\,\, U(\theta)=U(|\vec{p_1}-\vec{p}|)=  \nonumber \\
 U(\sqrt{p^2+p_1^2 - 2pp_1 \cos \theta}) 
\label{W}
\end{eqnarray}
is the even function of the scattering angle $\theta$. 
Indeed,  from  Eqs.(\ref{kinetic1},{\ref{St}) we obtain:
\begin{widetext}
\begin{eqnarray}
eE\frac{\partial f^{(0)}_+}{\partial p}= \frac{a_1p}{V_+(p)}f_{++}(p)-
\frac{p_+}{V_-(p_+)}(a_2 f_{++}(p) - a_3 f_{--}(p_+))
-m^2\tilde \alpha \int \frac{d\theta}{2\pi} \frac{dW(\theta)}{dp} \sin^2 (\frac{N\theta}{2})(1-\cos \theta)f_{++}(p)|_{\alpha=0}, 
\label{sm1} \\
eE\frac{\partial f^{(0)}_-}{\partial p} = \frac{a_1p}{V_-(p)}f_{--}(p)+ 
\frac{p_-}{V_+(p_-)}(a_3 f_{++}(p_-) - a_2 f_{--}(p)) 
+ m^2\tilde \alpha \int \frac{d\theta}{2\pi} \frac{dW(\theta)}{dp} \sin^2 (\frac{N\theta}{2})(1-\cos \theta)f_{++}(p)|_{\alpha=0},   
\label{sm2} 
\end{eqnarray}
 \end{widetext}
\begin{widetext}
\begin{eqnarray}
iN\frac{eE}{p}(f^0_+ - f^0_-) + \frac{i}{\hbar}(\epsilon_+ -\epsilon_- )(f_{+-}(p)+f_{-+}(p)) = 
+ia_4[\frac{p}{V_+(p)}f_{++}(p)- \frac{p}{V_-(p)}f_{--}(p)+  \nonumber \\
 \frac{p_-}{V_+(p_-)}f_{++}(p_-)- \frac{p_+}{V_-(p_+)}f_{--}(p_+)] 
 - im^2\tilde \alpha \int \frac{d\theta}{2\pi} \frac{dW(\theta)}{dp} \sin \theta \sin (N\theta)f_{++}(p)|_{\alpha=0}.
\label{sm3} 
\end{eqnarray} 
 \end{widetext}
In writing these equations we used Eqs.(\ref{D}). 
The coefficients $a_i$ in the above equations are given by the expressions
\begin{eqnarray}
a_1= \int \frac{d\theta}{2\pi} W(\theta) (\cos \theta-1)(1+\cos N\theta), \,\,  \nonumber \\
a_2= \int \frac{d\theta}{2\pi} W(\theta) (1-\cos N\theta), \nonumber \\
a_3= \int \frac{d\theta}{2\pi} W(\theta) \cos \theta (1-\cos N\theta), \,\, \nonumber \\
a_4= \int \frac{d\theta}{2\pi} W(\theta) \sin \theta \sin N\theta , \,\, \nonumber \\
a_5= \int \frac{d\theta}{2\pi} W(\theta) (1-\cos \theta).
\label{a} 
\end{eqnarray} 
Note that  here the scattering angle has an  arbitrary value and is not assumed to be small. 
 $f^0_+(p),f^0_-(p)$ are the equilibrium Fermi functions which correspond to the helicity 
$\pm$, $V_{\pm}(p)=p/m \pm N\tilde \alpha/2$ are the velocity values for a given $p$ for $\pm$ bands.
 The other quantities entering Eqs.(\ref{sm1}-\ref{sm3}) are defined in the following way
 \begin{eqnarray}
 p_{\pm}=p\pm m \tilde \alpha, \,\, V_-(p_+)=\frac{p}{m} - \frac{\tilde \alpha}{2}(N-2), \,\, \nonumber \\
  V_+(p_-)=\frac{p}{m} +\frac{\tilde \alpha}{2}(N-2),\,\,  \tilde \alpha= \alpha p_F^{N-1}.
\label{definition}
\end{eqnarray}
The quantities $V_-(p_+)$, $V_+(p_-)$ are the velocities in the bands $-$, $+$ for the momenta $p_+$, $p_-$, 
see Fig.1. These values of the momenta are connected in  Eqs.(\ref{sm1}-\ref{sm3}) because of the energy conservation under the elastic scattering. In writing Eqs.(\ref{definition}) we took into account that 
$m \tilde \alpha \ll p_F$, where $p_F$ is the Fermi momentum.

\begin{figure}
\begin{center}
\epsfig{file=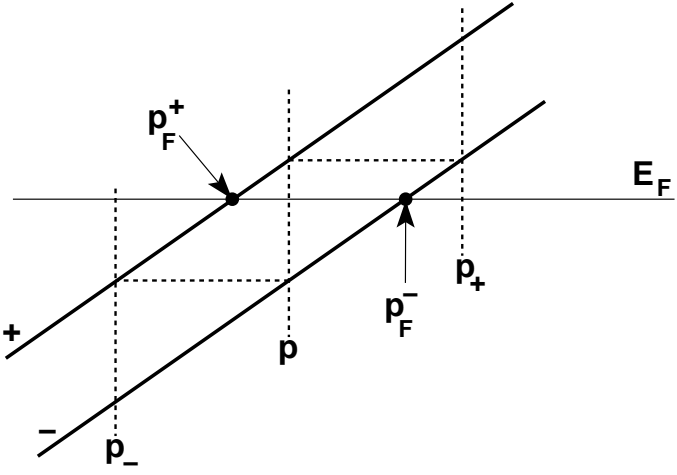,width=0.5\textwidth}
\end{center}
\caption{Schematics of the $\pm$ energy bands. Momenta $p, p_{\pm}, p^{\pm}_F$ are shown, see the text.}
\label{fig:SpinZero}
\end{figure}
 
 \par 
 Eqs.(\ref{sm1}-\ref{sm3}) are written for a given value of $p$. We see that the elastic scattering by impurities leads to admixture of the spin density matrix components which correspond to $p_+$, $p_-$ values of the momentum. 
 Different combinations of the ratio of the momentum and the velocity values like $p/V_+(p)$ or $p_+/V_-(p_+)$
 are just different density of state values corresponding to different points in the momentum space involved in the transitions, see Fig.1. From the form of the coefficients in Eq.(\ref{a}) we see that different terms in 
  Eqs.(\ref{sm1}-\ref{sm3}) have a simple physical meaning, namely, 
  they are just "transport" kinetic coefficients multipled (weighted) by the corresponding overlap factors 
  $\sin^2(N\theta/2)$, $\cos^2(N\theta/2)$ or $\sin(N\theta/2)\cos(N\theta/2)$  depending on 
  the type of  transition, i.e. whether it is intraband or interband scattering, see Eq.(\ref{D}) and Fig.1.   
 The very last term on the right hand side in each of the equations (\ref{sm1}-\ref{sm3})
 has a somewhat different origin. We can see that the scattering transitions within the band ($+$ or $-$) and between the bands correspond to different values of the $p_1$ momentum entering the kernel of the collision term, see Eq.(\ref{W}).
 In the case of the intraband transitions $p_1=p$, while for the interband transitions its value is either $p_+$ or 
 $p_-$. Thus, in the linear order with respect to $\tilde \alpha$  we obtain: 
 \begin{equation}
W(p \rightarrow p_{\pm})-W(q)=\pm \frac{dW(q)}{dp} \frac{m\tilde \alpha}{2}, W(q)=W(2p|\sin \frac{\theta}{2}|).
\label{W1}
\end{equation}
Note that the quantity $W(q)$ is the value of the scattering kernel for the case of intraband transition when 
the modulus of the momentum is conserved, $p_1=p$, and the modulus of the transferred momentum is $q=2p|\sin \theta/2|$. Hence, we have taken into account the correction to  $W(\theta)$ in the corresponding terms of Eqs.(\ref{sm1}-\ref{sm3}), keeping only the correction of the first order in $\alpha$, i.e. 
all the prefectors in the terms which already contain the derivative $dW(q)/dp$ are taken at $\alpha =0$. 
See, for example, the last term in Eq.(\ref{sm1}), where we also used that at $\alpha=0$ $f_{++}(p)=f_{--}(p)$ and given by Eq.(\ref{density}), see below. 
 
 \par
 All the quantities entering Eqs.(\ref{sm1}-\ref{sm3}) can be expressed through the corresponding values at $p$. 
 Writing  Eqs.(\ref{sm1}-\ref{sm2}) for the momenta values $p_{\pm}$, and using the relations 
  $ f^0_{+}(p)= f^0_{-}(p_+)$, $f^0_{+}(p_-)= f^0_{-}(p)$, we obtain 
\begin{eqnarray}
f_{--}(p_+)= f_{++}(p)\left[1-\frac{m\tilde \alpha NC}{Ap}\right] - \frac{m\tilde \alpha f_{++}(p)|_{\alpha=0}}{2A} 
\frac{da_1}{dp}- \nonumber \\
-\frac{\tilde \alpha \tilde E}{2Ap} (N-1), \nonumber \\
f_{++}(p_-)=f_{--}(p)\left[1+\frac{m\tilde \alpha NC}{Ap}\right] + \frac{m\tilde \alpha f_{++}(p)|_{\alpha=0}}{2A} 
\frac{da_1}{dp} + \nonumber \\
+\frac{\tilde \alpha \tilde E}{2Ap} (N-1), 
\label{f}
\end{eqnarray}
where $\tilde E$ is given by Eq.(\ref{density}) and $A,C$ are expressed through  coefficients $a_i$ by the formulas: $A=(a_1-a_2-a_3)/2, C=a_1+a_5$
 \begin{eqnarray}
 A= \int \frac{d\theta}{2\pi} W(\theta) (\cos \theta \cos N\theta -1), \nonumber \\  
 C= \int \frac{d\theta}{2\pi} W(\theta) (\cos \theta -1)\cos N\theta. 
 \label{A}
  \end{eqnarray}
 While deriving Eqs.(\ref{f}), we used the expansions
 \begin{eqnarray}
W(p_- \rightarrow p_-)=W(q)-m\tilde \alpha \frac{dW(q)}{dp}, \nonumber \\
W(p_+ \rightarrow p_+)= W(q)+ m\tilde \alpha \frac{dW(q)}{dp},
\label{W2}
\end{eqnarray}
where the right hand side of these equations is written for the momentum value $p$. Note the difference in the coefficients in the Eqs.(\ref{W2}) compared to Eq.(\ref{W1}). 
Using equation
  \begin{equation}
 [f_{++}(p) + f_{--}(p)]|_{\alpha=0}=-\frac{\tilde E}{m a_5}, \,\, \tilde E= eE \frac{\partial f_0}{\partial p},
 \label{density}
  \end{equation}
 which can be easily found from Eqs.(\ref{sm1},\ref{sm2}), with $f_0$ being the equilibrium Fermi function
 at $\alpha=0$, we find the final expression for the quantity in question, $(f_{+-}(p)+ f_{-+}(p))$, in the case 
 $m \tilde\alpha /\hbar \ll 1/R$:
  \begin{widetext}
 \begin{eqnarray}
 N\frac{eE}{p}(f^0_+ - f^0_-) + \frac{m\tilde \alpha \tilde E}{2a_5}  \left
[\frac{a_4}{A}\left (\frac{da_1}{dp} + \frac{2N}{p}a_2 -\frac{2(N-1)}{p}a_5 \right ) -\frac{da_4}{dp}\right ]
 = -\frac{(\epsilon_+ -\epsilon_- )}{\hbar}\left(f_{+-}(p)+f_{-+}(p)\right) 
\label{final}
  \end{eqnarray}
   \end{widetext}
 where $a_i$ are defined by Eq.(\ref{a}).
  Let us check some limiting cases from Eq.(\ref{final}).  
 \par
 1) We start with the simplest case of the $\delta$-correlated potential, when scattering is isotropic, i.e. 
 $W(\theta)$ is a constant. Then at $N\neq 1$ the coefficient $a_4=0$, and  from Eq.(\ref{final}) we immediately obtain a simple result: 
  \begin{equation}
 \frac{NeE}{p}(f^0_+ - f^0_-) - \frac{m\alpha \tilde E}{p} \delta_{N,1}
 = -\frac{(\epsilon_+ -\epsilon_- )}{\hbar}(f_{+-}(p)+f_{-+}(p)). 
  \label{deltascatt}
  \end{equation}
 With the help of Eqs.(\ref{q},\ref{angle2}) for the spin conductivity at $N\neq 1$  we obtain: 
 \begin{equation}
 \sigma_s =-\frac{eN}{8\pi \hbar}
  \label{short}
  \end{equation}
 Note that this result also follows trivially from the diagrammatic calculations since the vertex correction is identically zero. It happens because of a very simple reason, namely, the vector vertex contains the first harmonics but the Green's functions only the third one (at $N=3$), and their overlap is zero, as it happens here with the
  coefficient $a_4$.
 
 \par
 2) N=1. We know \cite{Bauer,Halperin,Khaetskii,Raimondi,Dimitrova,Chalaev} that in this case 
 for an {\it arbitrary} scattering (not nesessarely small angle or isotropic one) we should obtain  a zero value for $q_{yz}$. Indeed, in this case we have  $a_4=-A$, $a_1=-a_4$, $a_2=a_5$ and 
 from Eq.(\ref{final}) we easily obtain:
 \begin{equation}
 \frac{eE}{p}(f^0_+ - f^0_-) -\frac{m\alpha}{p} eE\frac{\partial f_0}{\partial p}
 = -\frac{(\epsilon_+ -\epsilon_- )}{\hbar}(f_{+-}(p)+f_{-+}(p)). 
  \label{N=1}
  \end{equation}
 Expanding the first function on the left hand side of this equation with respect to $\alpha$,  we obtain 
 a zero value for the spin current. 
 
 \par 3) Now consider the small-angle scattering case, $\theta \ll 1$. Then from Eq.(\ref{final}) it follows :
 \begin{eqnarray}
 \frac{NeE}{p}(f^0_+ - f^0_-) -\frac{m\tilde \alpha \tilde E}{p}\frac{2N(N^3-N+1)}{(N^2 +1)} - \nonumber \\
 N\frac{(N^2-1)}{(N^2 +1)} m\tilde \alpha \tilde E \frac{1}{a_5}\frac{da_5}{dp} 
 = -\frac{(\epsilon_+ -\epsilon_- )}{\hbar}(f_{+-}(p)+f_{-+}(p)). 
 \label{smallangle}
 \end{eqnarray}
 Finally we need to calculate the quantity $(1/a_5)(da_5/dp)$ in the limit of small-angle  scattering. 
 For this quantity we obtain 
 \begin{equation}
 \frac{1}{a_5}\frac{da_5}{dp}=\frac{1}{p}\frac{\int_0^{\infty} \frac{dq}{2\pi}\frac{\partial W(q)}{\partial q}\cdot q \cdot q^2}{\int_0^{\infty} \frac{dq}{2\pi} W(q) q^2}= -\frac{3}{p},
  \label{a5}
 \end{equation}
 where $q=p|\theta|$.  With the help of Eq.(\ref{a5}), we finally obtain for the spin conductivity 
 \begin{eqnarray}
 \sigma_s = -\frac{eN}{8\pi \hbar}[1+ \frac{(-5 +2N -2N^3+3N^2)}{(N^2 +1)}]= \nonumber \\
 = \frac{eN}{4\pi \hbar}\frac{(N^2-1)(N-2)}{(N^2 +1)}.
 \label{sigma}
 \end{eqnarray}
The first term (unity) inside the square brackets originated from the first term on the left hand side of Eq.(\ref{smallangle}). The result we obtain, i.e. Eq.(\ref{sigma}), coincides exactly with the corresponding result of Ref.\onlinecite{Shytov2}.

\par
4) Finally, we consider the case of a very smooth scattering potential when the impurity radius obeys the inequality
$1/R \ll m\tilde \alpha/\hbar \ll k_F$. \cite{Khaetskii1} Physically it means that elastic transitions between different bands (with opposite helicity) are forbidden since the quantity $W(\theta)$, see Eq.(\ref{W}), is exponentially small for these transitions. Then in Eqs.(\ref{sm1}-\ref{sm3}) we should drop all the terms containing $p_{\pm}$ quantities (and also derivatives $dW/dp$), and with the use of the fact, that in the case considered $\theta \ll 1$,  
the final result reads:
\begin{eqnarray}
 \frac{eEN}{p}(f^0_+ - f^0_-) + NeE(\frac{\partial f_+^0}{\partial p}-\frac{\partial f_-^0}{\partial p})
 = \nonumber \\
 -\frac{(\epsilon_+ -\epsilon_- )}{\hbar}(f_{+-}(p)+f_{-+}(p)). 
 \label{verysmooth}
 \end{eqnarray}
Integrating in Eq.(\ref{q}) between the points $p_F^{\pm}= \mp m\tilde \alpha/2 + p_F$, see Fig.1, we obtain for the spin conductivity 
\begin{equation}
 \sigma_s =-\frac{eN}{8\pi \hbar}(N-1). 
  \label{verysmooth1}
  \end{equation}
 which for $N=3$ is twice as large compared to the $\delta$-scattering case. \cite{Khaetskii1}

\par
In conclusion, we have studied how intrinsic spin current depends on the form of the Hamiltonian and the scattering potential properties.
As an example, we have investigated the generalized 2D Rashba model when the spin-orbit term contains an arbitrary dependence on the electron momentum. We have found an exact analytic solution for the intrinsic spin current 
 in the ballistic limit, when spin-orbit splitting is much larger than the disorder induced broadening. In contrast to the case of a common, linear in momentum Hamiltonian, the intrinsic  current does not necessarily vanish.  However, even in the ballistic limit indicated above, its value depends  explicitely on the disorder properties. More precisely, the result, being independent on the spin-orbit coupling constant, is different for a different correlation radius of the disorder potential.
  The situation is similar for the other Hamiltonians. For example, we have also studied this problem for the Hamiltonian of 3D holes. \cite{Luttinger} 
 In the diagrammatic language it means that even in the ballistic limit the vertex correction is always as important as the contribution of the empty loop and 
 physically it is not correct to consider them separately. 
 In this sense, universal intrinsic spin current does not exist. 

\par
I am grateful to A.H. MacDonald and T. Jungwirth for fruitful discussions.
I am also grateful to A. Shytov,  E. Mishchenko,  B.I. Halperin and H.A. Engel, 
discussions with whom helped to correct errors in the previous versions of
our papers \cite{Shytov,Khaetskii2}.



\begin{references}
\bibitem{Luttinger}
We have done similar calculations for the holes described by the Luttinger Hamiltonian, A. Khaetskii, unpublished. 
The calculations are done within the Born approximation for the collision integral when the {\it extrinsic} spin current is zero \cite{Dyakonov1}.
 These calculations show that the conclusion of the paper by S. Murakami, Phys. Rev.B {\bf 69}, 241202(R) (2004), about the complete absence  of the corrections to the intrinsic value of the spin current  due to scattering is correct only for the $\delta$-correlated potential.
\end{references}
\end{document}